\documentclass[aps,prd,twocolumn,preprintnumbers,nofootinbib,floatfix,reprint]{revtex4-1}

\usepackage{dcolumn}
\usepackage{graphicx}
\usepackage{amsmath}
\usepackage{amssymb}
\usepackage{amsthm}
\usepackage{latexsym}
\usepackage{color}
\usepackage{bm}
\usepackage{multirow}
\usepackage{cancel}
\usepackage{hyperref}
\usepackage{booktabs}

\newcommand*{\beq}{\begin{equation}}
\newcommand*{\eeq}{\end{equation}}
\newcommand*{\bea}{\begin{eqnarray}}
\newcommand*{\eea}{\end{eqnarray}}

\newcommand{\Lag}{\mathcal L}

\begin{document}
 
\title{The 750 GeV threshold to a new particle world }
 
\author{Samarth Jain$^1$, Fabrizio Margaroli$^{2,3}$, Stefano Moretti$^{1,3}$ and Luca Panizzi$^{1,3}$} 

\affiliation{$^1$ School of Physics \& Astronomy, University of Southampton, Highfield, Southampton SO17 1BJ, UK\\
$^2$ Department of Physics, Sapienza University Rome, Edificio G. Marconi, 00185 Roma, Italy \\
$^3$ Department of Physics, Rutherford Appleton Laboratory, Chilton, Didcot  OX11 0QX, UK}

\date{\today}

\begin{abstract} 
We show how an excess in the diphoton channel can be the effect of neither a resonance nor an end-point in a cascade decay, but rather of a threshold for virtual production of a pair of extra quarks, each with half of peak invariant mass, onsetting in both the $gg$-initiated production and the $\gamma\gamma$-induced decay of an off-shell $Z$ boson. For our analysis we consider as paradigmatic example the 750 GeV excess previously seen at the end of 2015 with the Run 2 data of the LHC but not confirmed with 2016 data.
\end{abstract}


\maketitle


A  bump appeared in 2015 at around 750 GeV in the diphoton invariant mass distribution extracted at the 13 TeV run of the CERN Large Hadron Collider (LHC) in both ATLAS and CMS analyses. 
The excess was observed with a local significance around 3.9$\sigma$ by ATLAS~\cite{750GeV-ATLAS} using 3.2 fb$^{-1}$ data and around 3.4$\sigma$ by CMS~\cite{CMS:2015dxe} using 3.3 fb$^{-1}$ data. The excess had no obvious characteristic shape. In fact, innumerable publications appeared in the literature attempting to explain is as either a $s$-channel Breit-Wigner resonance (typically a (pseudo)scalar one) decaying into $\gamma\gamma$ or as an end-point Jacobian shoulder (typically induced by a cascade decay of an heavy object producing two photons alongside some amount of missing energy). In fact, the quality of the ATLAS and CMS data around the excess was not sufficient to separate the two hypotheses. 

In the light of this, we attempt here an alternative explanation inspired by the work in Ref.~\cite{Moretti:2014rka}, which exploits a threshold effect induced by the creation of a virtual pair of new extra quarks in a loop diagram. We stress however that this analysis can be applied to any excess in the diphoton channel that might be observed in the future, with suitable numerical adaptations. But let us proceed in steps. First, notice that, contrary to popular belief, a $gg$ pair can annihilate into a massive spin-1 object which can also in turn decay into a $\gamma\gamma$ pair. This is possible so long that the intermediate (massive) gauge boson is off-shell, otherwise the Landau-Yang theorem \cite{Landau:1948kw,Yang:1950rg} would prevent this possibility. Secondly, we argue here (as in Ref.~\cite{Moretti:2014rka}) that this can be an off-shell $Z$-boson (hereafter, denoted by $Z^*$). In this case, the only (non-resonant) component of the $Z$ boson that is allowed to propagate in this channel is its Goldstone one (this phenomenon is indeed best appreciated in the Landau gauge, $\xi\to0$), which is CP-odd. 
Hence, both $gg\to Z^*$ production and $Z^*\to\gamma\gamma$ decay\footnote{Notice that the language adopted here is merely for illustration purposes, as we adopt here the same approach of Ref.~\cite{Moretti:2014rka}, which computed the full process $gg\to Z^*\to\gamma\gamma$, without any factorisation.} 
need to be mediated by loops of fermions.
Thirdly, we conclude our interpretation by invoking, alongside the SM quarks propagating in both loops (specifically, quarks only in production and both  
quarks and leptons in decay, though the dominant contribution is always the one due to the top quark with $m_t=173$ GeV), also the presence of an additional extra quark with mass 375 GeV. Hence, the threshold that appears in the calculation as the imaginary part of each loop when $m_{gg}\equiv m_{\gamma\gamma}= 2 m$, and manifests itself as a local enhancement at such a mass value, could well be
responsible for the aforementioned excess at 750 GeV. Other analyses have appeared in literature considering threshold effects to explain the excess. However usually either new particles propagating in the s-channel\cite{Djouadi:2016eyy,Bharucha:2016jyr,DiChiara:2016dez} or resonant bound states have been invoked\cite{Luo:2015yio,Han:2016pab,Kats:2016kuz}. A scenario analogous to the one we will analyse, with just one new particle propagating in the loops, has been considered in Ref.\cite{Chway:2015lzg}; however, in that analysis a larger number of free parameters (mass, couplings and width of the heavy quark) have to be fixed to fit the data.

Of course, not any extra quark can be responsible for such a phenomenology. On the one hand, it should have an axial 
coupling to the aforementioned $Z^*$ boson (i.e., to its Goldstone component in a general $R_\xi$ gauge),
otherwise its contribution to the whole $gg\to Z^*\to\gamma\gamma$ process would vanish identically by Furry's theorem \cite{Furry:1937zz,Dyson:1949ha,Feynman:1949hz}.  On the other hand, following the discovery of a Higgs boson \cite{Aad:2012tfa,Chatrchyan:2012xdj} with essentially a Standard Model (SM) nature, the existence of an additional chiral quark (i.e., with SM-like $V-A$ structure in gauge boson charged currents) has been excluded \cite{Djouadi:2012ae,Eberhardt:2012gv}.
Vector-Like Quarks (VLQs) seem to perfectly fit this bill. These hypothesised states of matter are heavy spin 1/2 particles that transform as triplets under colour but, unlike SM quarks, their left- and right-handed couplings have the same Electro-Weak (EW) quantum numbers.  

The ATLAS and CMS collaborations have carried out a broad program of searches for VLQs with different quantum numbers, probing single and pair production mechanisms, as well as decay modes into all three generations of SM quarks (for the most updated experimental results of ATLAS and CMS we refer to the respective web pages \cite{twikiATLAS8TeV,twikiATLAS13TeV,twikiCMS}). However, new extra quarks can be charged under new symmetries, as T-parity in Little Higgs \cite{ArkaniHamed:2002qx,Cheng:2003ju,Cheng:2004yc,Low:2004xc,Hubisz:2004ft,Cheng:2005as,Hubisz:2005tx} and Kaluza-Klein parity in Extra Dimensions (EDs)  
\cite{Antoniadis:1990ew,Appelquist:2000nn,Servant:2002aq,Csaki:2003sh,Cacciapaglia:2009pa}, thus forbidding the decay to SM particles but allowing decays to Dark Matter (DM) candidates instead. Such VLQs have been searched for  at both the Tevatron \cite{Aaltonen:2011rr, Aaltonen:2011na} and LHC \cite{ATLAS:2011mda,CMS:2012dwa}, but the constraints on their mass and decays depend on the mass of the DM candidate and on which SM quarks they decay to. Specifically, if the VLQ and the DM candidate have a strong degeneracy, the visible decay products of the VLQ are too soft to be detected and, as a consequence, the bounds on the VLQ masses can be very weak, analogously to the case of strong degeneracy between squarks and neutralinos in supersymmetry. 

\begin{figure}[!t]
\centering
\begin{minipage}[c]{.26\textwidth}
\includegraphics[width=\textwidth]{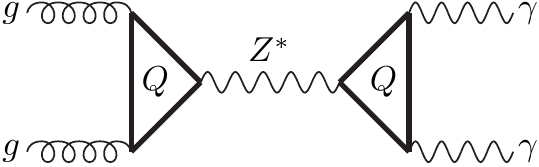}
\end{minipage}\hfill
\begin{minipage}[c]{.19\textwidth}
\includegraphics[width=\textwidth]{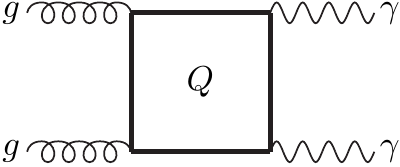}
\end{minipage}
\caption{Gluon-gluon induced diphoton production through $Z^*$ exchange and a Box. Herein, $Q$ refers to a generic massive quark. The lepton component in the $Z^*\gamma\gamma$ loop is negligible.}
\label{fig:Feynman}
\end{figure}

It is the purpose of this letter to compute the $gg\to Z^*\to\gamma\gamma$ process in the SM supplemented by a VLQ with mass $m_{\rm VLQ}=375$ GeV (\textit{i.e.} half of the invariant mass value of the excess), odd under a $\mathcal Z_2$ symmetry under which SM particles are even, charge 2/3$e$ - a vector-like top quark partner - and arbitrary coupling strength to the $Z^*$ boson and compare its yield to the aforementioned ATLAS and CMS data displaying the described 750 GeV excess. In fact, for completeness (although its contribution is subleading with respect to $Z^*$ mediation), we also include an estimate of the $gg\to{\rm Box}\to \gamma\gamma$ channel, which also displays a threshold at $2m_{\rm VLQ}$. The corresponding Feynman diagrams are in Fig.~\ref{fig:Feynman}.

The generic interaction Lagrangian of a a singlet VLQ $Q$ coupled to the SM $Z$ boson is written as:
\begin{eqnarray}
\Lag_{Q,Z} &=& \bar{Q} \gamma_\mu \left(g^{QZ}_L P_L + g^{QZ}_R P_R \right) Q Z^\mu, 
\end{eqnarray}
where $g^{QZ}_{L/R}$ are the left/right-handed couplings of the vector-like quark $Q$ to the $Z$ boson and $P_{L/R}$ the chiral projection operators. The couplings of the VLQ to the photon and gluon are SM-like.
In what follows, we will perform the numerical analysis for the case of a single VLQ with mass $m_{\rm VLQ}=375$ GeV which couples to (the Goldstone component of) the $Z$ boson $\propto g_Z\mu m_{\rm VLQ}/M_Z$\footnote{We can extract the mass from the coupling, e.g. assuming that the VLQ singlet also gets a Dirac mass term by interacting with the Higgs and a much heavier VLQ doublet, such that the masses of the physical states are obtained through a see-saw mechanism.}, where $g_Z=e/(2\sin\theta_W\cos\theta_W)$ and $\mu$ is a rescaling factor of the EW interaction strength which characterises
the VLQ versus $Z$ coupling. When considering theoretically motivated scenarios, in general $\mu\simeq 1$: the coupling between the VLQ and the SM $Z$ boson is usually of EW strength, but it can be corrected by mixing angles which parametrise rotations in the gauge and/or VLQ sectors (these corrections, being induced through mixing angles, are necessarily smaller than 1) or EW loop corrections, which are small, usually of the order of $(4 \pi)^{-1}$. The masses of the VLQs are also usually larger than the value we are considering, as these states are related, e.g., to the radius of the EDs or to the scale of EW symmetry breaking and are further constrained by complementary observables as relic density or EW precision tests. However, for the purpose of our analysis and in order to be as model-independent as possible, we will consider $\mu$ as a free parameter and fix the mass of the VLQ to be 375 GeV to (approximately) fit the excess in the data. Needless to say, by browsing the appendix of Ref.~\cite{Moretti:2014rka}, it is clear that the same result obtained for a single VLQ with a $\mu$ coupling rescaling can be obtained from a family of (nearly) degenerate VLQs of $n$ members each a with $\mu/n$ like-sign modification (without anomaly cancellations as, being $\mathcal Z_2$-odd, they do not mix with SM states), so that our illustrative results can be mapped in whatever realistic VLQ scenario is required.
Further, notice that, for the case of the $gg\to {\rm Box}\to \gamma\gamma$ contribution, being the VLQ couplings to gluons and photons fixed by their QCD and QED charges, the overall VLQ dependence at the amplitude level is $n$ times the one of a single VLQ with
mass 375 GeV. Between the two types of contributions, though, for $m_{\gamma\gamma}$ around 750 GeV, we will show that it is the $Z$ one to dominate over the Box one (so that their interferences is also negligible), primarily because the former, being a two-loop topology, sees an overall VLQ contribution at amplitude level which scales as $n^2$ whereas the latter, being a one loop-topology, only scales as $n$. {This also means that the $gg\to Z^*\to \gamma\gamma$ channel displays a characteristic $\mu^4$ dependence at threshold.}

\begin{figure}[!t]
\centering
\includegraphics[height=0.95\linewidth,angle=90]{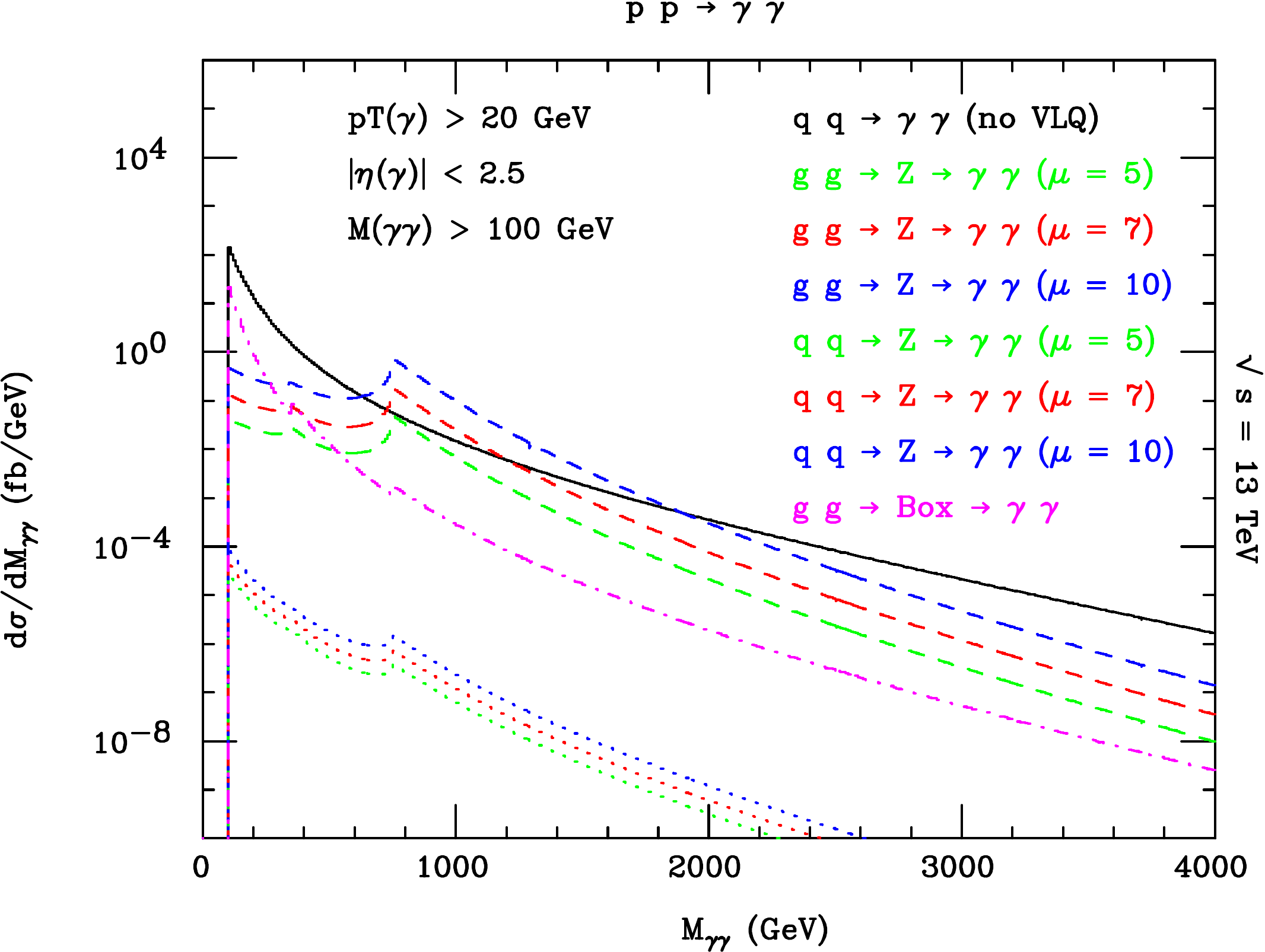}
\caption{Differential diphoton mass distributions at the 13 TeV LHC for the processes 
$q\bar q\to \gamma\gamma$ (solid),
$gg\to Z^*\to \gamma\gamma$ (dashed),
$q\bar q\to Z^*\to \gamma\gamma$ (dotted) and
$gg\to {\rm Box}\to \gamma\gamma$ (dot-dashed) after the cuts
$p^T_\gamma>20$ GeV, $|\eta_\gamma|<2.5$ and $M_{\gamma\gamma}>100$ GeV.
The three lines for the $Z^*$  induced processes are for $\mu=5,7$ and 10 (from bottom to top).
CTEQ(5L) with $Q=\mu=\sqrt{\hat s}$ is used \cite{Lai:1999wy}.}
\label{fig:Fabrizio}
\end{figure}

With the model put  forward, we proceed to investigate the possibility that the diphoton excess observed in the ATLAS and CMS experiments could have been indeed due to the threshold production via a $Z^*$ of new particles of mass approximately 375\,GeV. The ATLAS collaboration designed the diphoton analysis with two slightly different event selections optimised for the search of a spin-0 or a spin-2 boson. We use in this paper the former, but expect the latter to yield equivalent results. The CMS collaboration uses the same analysis for interpretations in terms of spin-0 and spin-2 resonances. However, CMS splits the data into four subsamples depending on whether both photons hit the EM Barrel (EB) calorimeter, located in the central region - EBEB events - or one hits the EB calorimeter and the other one the EM Endcap (EE) calorimeter  located in the forward region - EBEE events. The data are further split into events where the magnet is either on at full capability, 3.8T, or off, 0T. We extract the data from LHC publications and keep the same subsample division while binning in invariant mass of the two photons. We compute the differential  cross section as a function of the diphoton invariant mass for the $gg \rightarrow Z^* \rightarrow \gamma \gamma$ and $gg \rightarrow {\rm Box} \rightarrow \gamma \gamma$ processes, both mediated by loops of fermions including the VLQs, and the SM $q \bar q \rightarrow \gamma \gamma$ tree-level process, as shown in Fig.~\ref{fig:Fabrizio}. The process mediated by the $Z^*$ and the background contributions have been computed with a fortran code based on the HELAS libraries~\cite{Stelzer:1994ta} and considering the {\sc CTEQ(5L)} PDF set~\cite{Lai:1999wy}, while for the box contribution, due to its smallness in the relevant region, the curve has been computed by considering loops with massless quarks and approximating the top and VLQ contributions with step functions at the correponding mass thresholds.
It is clear that the dominant contribution at $m_{\gamma\gamma}=750$ GeV is due to the $gg \rightarrow Z^* \rightarrow \gamma \gamma$ process. 

The signal distribution displays two characteristic non-resonant peaks due to the thresholds induced by the heavy SM and additional quark. The first peak at approximately 350\,GeV is caused by the loop of top quarks and is not visible due to the background process yielding at least one order of magnitude more events. The threshold at around 750\,GeV is induced by the loop of VLQs and appears as a peak with a long tail extending to the right side of the distribution. 
We fit the virtual VLQ pair production signal using a Landau function on top of an exponentially falling background distribution and we checked the dependence of the shape at threshold
as a function of the choice of the EW couplings of the VLQ and found negligible differences over the range of $\mu$ values explored in this paper.  We use the same experimental event selection efficiencies and acceptances the ATLAS and CMS computed for a spin-0 resonance, to model event selection efficiency times acceptance of our signal. 
The photon momentum spectrum and pseudo-rapidities in our threshold production scenario are expected to be approximately similar to the ones induced by Breit-Wigner resonances explored in the LHC papers. Also, we use the CMS parametric choice to fit the falling background diphoton spectrum.

\begin{figure}
\centering
\includegraphics[width=0.74\linewidth]{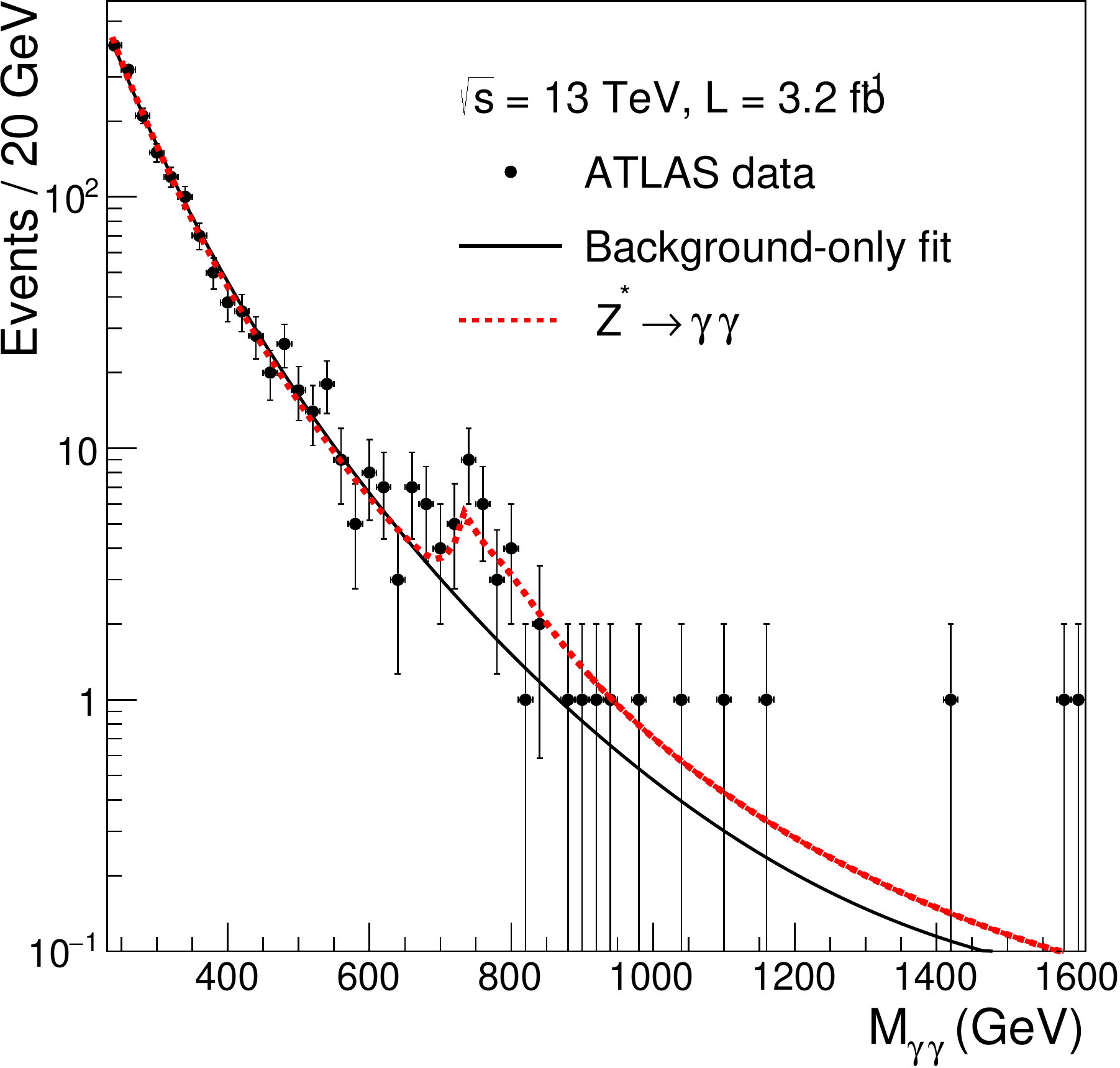}
\caption{The plot shows the ATLAS diphoton invariant mass spectrum after the event selection in \,\cite{750GeV-ATLAS}. The black solid curve shows the background-only fit to the data, while the red dashed curve shows the result of the fit for the background plus non-resonant signal hypothesis.} 
\label{fig:ATLAS}
\end{figure}

\begin{figure}
\centering
\includegraphics[width=0.95\linewidth]{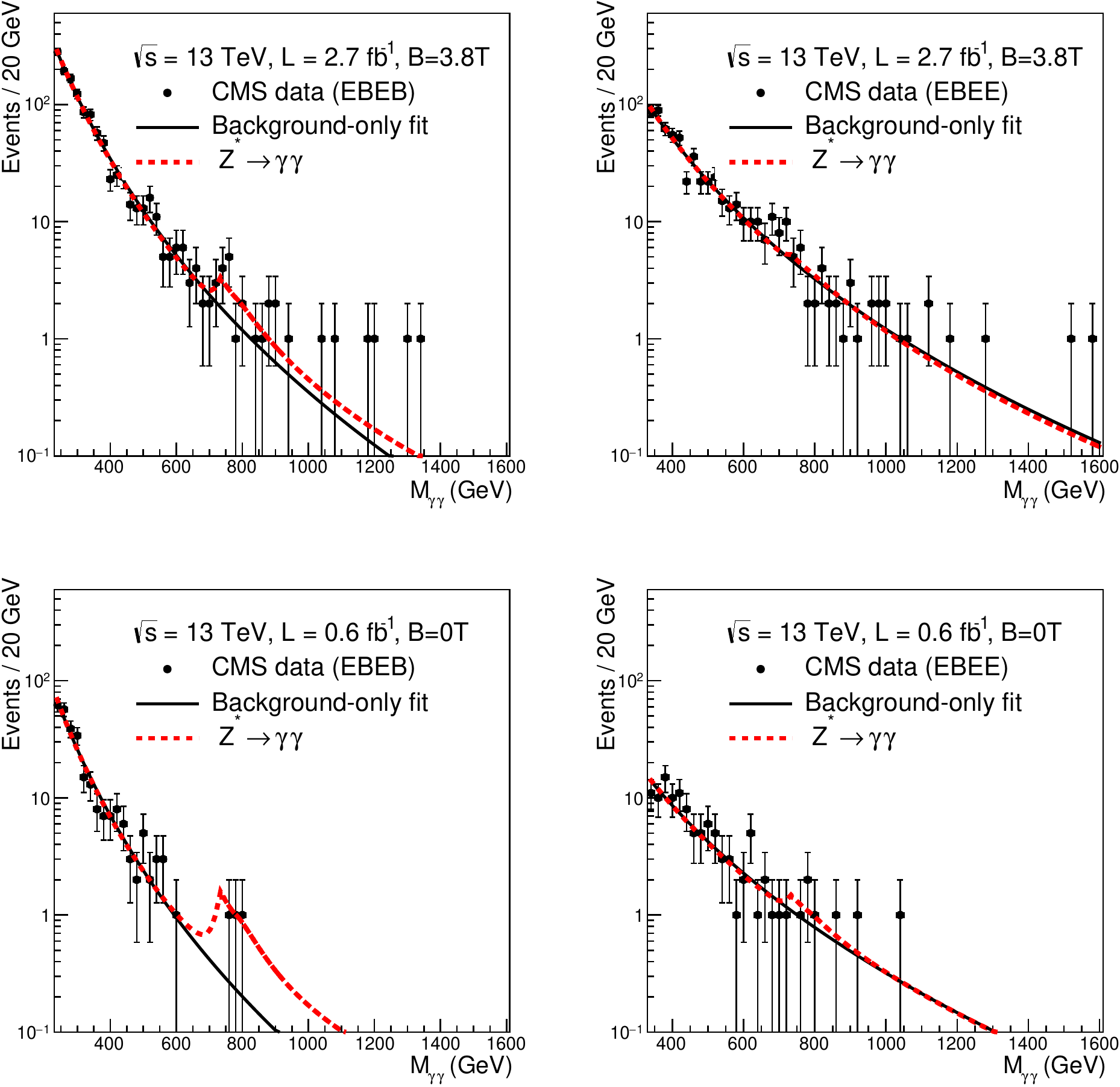}
\caption{Same as Fig.\ref{fig:ATLAS} but for the CMS diphoton invariant mass spectrum after the event selection in \,\cite{CMS:2015dxe} divided in the four subsamples depending on the appearance in the detector of the two photons. The red curve is normalized to the $\mu$ value measured simultaneously in the four subsamples.} 
\label{fig:CMS}
\end{figure}

The data generally prefer the presence of a signal, and regard similarly the resonant and non-resonant hypothesis. 
Considering a VLQ mass of 375 GeV the goodness of the fit ($\chi^2$) for ATLAS is mildly better in the case of a Breit-Wigner (BW) rather than the hypothesis considered in this paper, while for CMS the threshold hypothesis is sligthly favoured. Considering instead a VLQ mass of 365 GeV, for ATLAS the threshold hypothesis has a comparable $\chi^2$ to the BW case, while for CMS the goodness of the fit improves in two subsamples but becomes worse in the remaining two, as it can be seen in Tab.~\ref{tab:chi2}.
\begin{table}
\setlength{\tabcolsep}{5pt}
\begin{tabular}{ccccc}
\multicolumn{2}{c}{signal hypothesis} & \multicolumn{2}{c}{Threshold} & BW \\
\multicolumn{2}{c}{$m_{\rm VLQ}$}     & 365 GeV & 375 GeV & 375 GeV \\
\hline
\multicolumn{2}{c}{ATLAS}             & 0.658   & 0.729   & 0.651 \\
                                                           
\hline                                                     
\multirow{4}{*}{CMS} & EBEB 3.8T       & 0.753   & 0.765   & 0.755 \\
                     & EBEE 3.8T       & 0.875   & 0.853   & 0.866 \\
                     & EBEB 0T         & 0.756   & 0.763   & 0.887 \\
                     & EBEE 0T         & 0.550   & 0.536   & 0.586 \\
\hline
\end{tabular}
\caption{\label{tab:chi2} Values of the $\chi^2$ for both experiments and for different values of the VLQ mass in the Breit-Wigner and threshold hypotheses.}
\end{table}

Figs.~\ref{fig:ATLAS} and \,\ref{fig:CMS} show the result of the fit according to two  different hypotheses: the background-only hypothesis, where the background is parametrised as in \cite{CMS:2015dxe}, and the threshold production for VLQs. We use the signal normalisations extracted from the fits to the different event selections and the corresponding efficiencies and acceptances computed by the ATLAS and CMS collaborations to extract a measurement of the EW coupling of the VLQ to the $Z$ boson. The ATLAS data favour a rescaling factor $\mu$ of the EW coupling of 10.5 while the CMS ones point to $\mu$=10.9. 

These couplings are admittedly quite large, possibly already in the non-perturbative region, and the reinterpretation of our results in terms of a theoretical model is not straightforward. However, as mentioned above, our results can be obtained either by keeping the coupling as a free parameter or by fixing the coupling and including degenerate VLQ copies. The inclusion of more VLQs would be also theoretically justified by requiring each SM quark to have a $\mathcal Z_2$-odd partner, as in the case of universal EDs. As an example, if we allow the number of degenerate VLQs to be a free parameter, and we fix it to 12 (6 top-like and 6 bottom-like, i.e., a partner for each chirality of SM quarks) we can fit the ATLAS data with a value of $\mu={10.5\over 6 + 6 (-1/3)^4/(2/3)^4 }\simeq1.6$ and the CMS ones with $\mu\simeq1.7$. 

It is interesting to note that the local significances observed by the ATLAS and CMS collaborations of respectively 3.9$\sigma$ and 3.4$\sigma$ quoted in the context of the Breit-Wigner hypothesis drop to approximately $2.5\sigma$ and $2.2\sigma$ when assuming the VLQ threshold production hypothesis. If the excess is real, but not of resonant nature, the combined LHC sensitivity using existing data would be approximately $3\sigma$ and more data would be needed for an unambiguous confirmation of the current excess. The LHC experiments could easily adjust the analysis strategy to encompass the possibility highlighted in this paper. In particular, the ATLAS collaboration already fits the background using both a data-driven approach and a complete simulation of the backgrounds. Instead, 
the CMS experiment uses a data-driven approach, thus the prediction at the high-end tail of the diphoton spectrum relies on the limited statistics acquired as of now.

To conclude, it is very interesting to entertain the possibility that a new and unexpected signal could have been present in the LHC data. However, the vast literature that surfaced after the first appearance of the 750\,GeV excess suggests that there is a large number of possibilities to explore, mostly requiring a zoo of new particles of varied nature to build a consistent framework. We have instead put forward in this paper a particularly economical  solution, already intimated by Ref.~\cite{Moretti:2014rka}, that requires only the existence of a new quark of vector-like nature (VLQ), or a small set of near-degenerate such states, with mass approximately 375\,GeV. The 750\,GeV bump would have been therefore the result of the threshold production of one or more (degenerate) pairs of such (virtual) particles. We have shown the the LHC data were consistent with this new hypothesis and we extracted the EW coupling between the VLQ and the $Z^*$ boson and/or the number of generations of VLQs that would be needed to describe the data. Finally, we remark that a by-product of our scheme, due to the fact that it is actually the Goldstone component of the $Z^*$ state that intervenes in our case, is that its pseudoscalar {\sl tree-level} properties conveniently explain why no excess has been found yet in either the $W^+W^−$ or $ZZ$ data sample \cite{Sato:2016hls,No:2016htu}. Conversely, because of the intense $Z-{\rm VLQ}$ interaction, sensitivity to $Z\gamma$ final states is also potentially possible.\\

\section*{Note added}
During the  ICHEP 2016 conference, hence after the submission of this work to the journal,
both ATLAS and CMS updated their  earlier results by employing
an increased data sample 
based on  some 12 fb$^{-1}$ of LHC Run 2 data per experiment: see the talks by
B. Lenzi (ATLAS) and  C. Rovelli (CMS) at
{\tt https://indico.cern.ch/event/432527/contributio}\\
{\tt ns/1072431/}. The new LHC results around 750 GeV are now compatible with the SM at the level of less than 
$2\sigma$. Therefore, data in the $\sim750$ GeV region are, at present, consistent with a statistical 
fluctuation. However, it remains of importance to explore the high mass region for two reasons.
Firstly, the fact that both ATLAS and CMS initially recorded a very similar excess calls for a closer scrutiny than normal of 
data  as more and more luminosity will accrue.
Secondly, a coordinated effort between theorists and 
experimentalists  is presumably required in order to carefully address all possibilities explaining potential anomalies,
from the more evident ones (like resonances) to the more subtle ones (like those studied here).

\begin{figure}
\centering
\includegraphics[height=0.9\linewidth,angle=90]{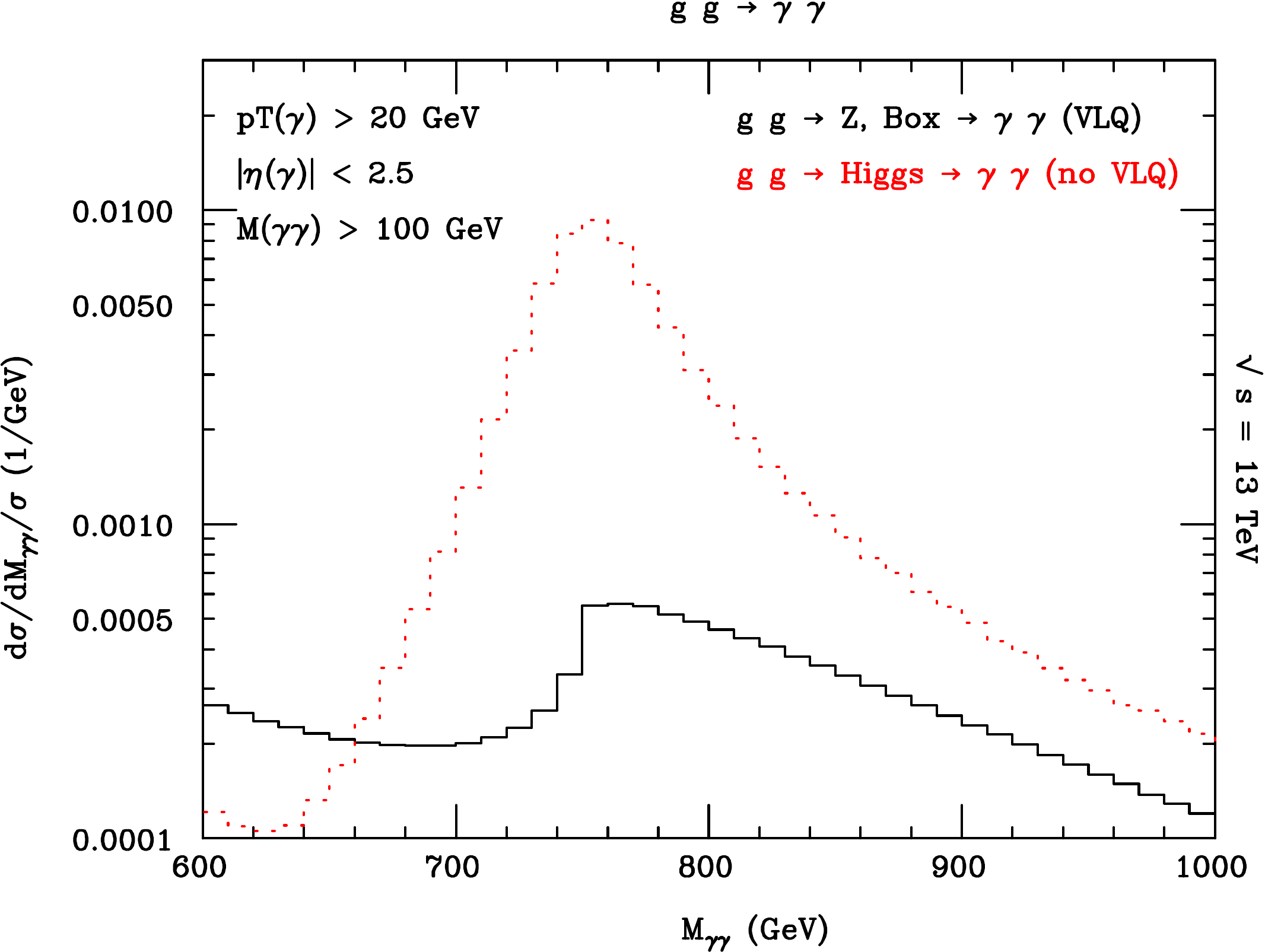}
\caption{Normalised (to 1) differential diphoton mass distributions at the 13 TeV LHC for the processes 
$q\bar q\to Z^*, {\rm Box}\to \gamma\gamma$ (solid),  with $m_{\rm VLQ}=375$ GeV and $\mu=1$
(VLQ induced), as well as 
$gg\to {\rm Higgs}\to \gamma\gamma$ (dotted), with $M_{\rm Higgs}=750$ GeV, $\Gamma_{\rm Higgs}=50$ GeV and
SM-like couplings (CP-even scalar mediated), after the cuts
$p^T_\gamma>20$ GeV, $|\eta_\gamma|<2.5$ and $M_{\gamma\gamma}>100$ GeV.
CTEQ(5L) with $Q=\mu=\sqrt{\hat s}$ is used \cite{Lai:1999wy}.}
\label{fig:BW-Threshold}
\end{figure}

But let us first  recap the  situation. Tab.~\ref{tab:ICHEP} does so  in statistical terms before and after 
the ICHEP 2016 conference. It should also be noted that, while CMS retains a moderate excess
at 750 GeV,  ATLAS now favours a mass value slightly lower (730 GeV) or higher (770 GeV) than previously. In short,
the local significance is still slightly pronounced (at the $2\sigma$ level) while we estimate the global one to be rather 
poor (at the $1\sigma$ level). However, two considerations are in order. On the one hand, 
notice that, with twice as much luminosity as the present one, it would be possible
to re-obtain a $\approx3\sigma$ significance per experiment. On the other hand, it should be appreciated that 
only the $\sigma\times$ BR hypothesis has been tested. Hence, the dynamics we propose, i.e., a less pronounced threshold enhancement (which is approximately modelled by an asymmetric Landau distribution)  than a resonant one (which is approximately modelled by a BW), as manifest from Fig.~\ref{fig:BW-Threshold}, remains untested to this day. Unfortunately, with data currently unavailable, we are not in the position to test it ourselves. We look forward to the LHC experiments to tackle this challenge  rapidly.

\begin{table}
\setlength{\tabcolsep}{5pt}
\begin{tabular}{c||c|c}
\hline
{} & {ATLAS} & {CMS} \\
\hline\multicolumn{3}{c}{}\\[-8pt]
\multicolumn{3}{c} {~~Pre-ICHEP} \\[2pt]
\hline
local $p$-value                     &        $3.9\sigma$ & $3.4\sigma$\\
limit on $\sigma\times$~BR, NWA & $\approx12$ fb & $\approx 14$ fb\\
fitted $\sigma\times$~BR, NWA                &     $\approx8$ fb & $\approx 4$ fb\\
\hline\multicolumn{3}{c}{}\\[-8pt]
\multicolumn{3}{c} {~~Post-ICHEP} \\[2pt]
\hline
local $p$-value                       &      $2\sigma$ & $2\sigma$\\
limit on $\sigma\times$~BR, NWA(wide) & $\approx2$ fb & $\approx 2(4)$ fb\\
fitted $\sigma\times$~BR, NWA(wide)                &     $\approx1$ fb & $\approx 1(2)$ fb\\
\hline
\end{tabular}
\caption{\label{tab:ICHEP} ATLAS and CMS statistical findings before and after ICHEP.}
\end{table}

\vspace*{0.25cm}
\noindent
{\emph{Acknowledgements}} SM and LP are supported in part through the NExT Institute  and the STFC Consolidated Grant ST/J000396/1. FM is supported in part through the {\em Rita Levi Montalcini 2009 - MIUR} grant. FM wishes to thank Chiara Rovelli for the useful discussions.

\bibliographystyle{apsrev4-1}
\bibliography{XF.bib}

\begin{thebibliography}{42}%
\makeatletter
\providecommand \@ifxundefined [1]{%
 \@ifx{#1\undefined}
}%
\providecommand \@ifnum [1]{%
 \ifnum #1\expandafter \@firstoftwo
 \else \expandafter \@secondoftwo
 \fi
}%
\providecommand \@ifx [1]{%
 \ifx #1\expandafter \@firstoftwo
 \else \expandafter \@secondoftwo
 \fi
}%
\providecommand \natexlab [1]{#1}%
\providecommand \enquote  [1]{``#1''}%
\providecommand \bibnamefont  [1]{#1}%
\providecommand \bibfnamefont [1]{#1}%
\providecommand \citenamefont [1]{#1}%
\providecommand \href@noop [0]{\@secondoftwo}%
\providecommand \href [0]{\begingroup \@sanitize@url \@href}%
\providecommand \@href[1]{\@@startlink{#1}\@@href}%
\providecommand \@@href[1]{\endgroup#1\@@endlink}%
\providecommand \@sanitize@url [0]{\catcode `\\12\catcode `\$12\catcode
  `\&12\catcode `\#12\catcode `\^12\catcode `\_12\catcode `\%12\relax}%
\providecommand \@@startlink[1]{}%
\providecommand \@@endlink[0]{}%
\providecommand \url  [0]{\begingroup\@sanitize@url \@url }%
\providecommand \@url [1]{\endgroup\@href {#1}{\urlprefix }}%
\providecommand \urlprefix  [0]{URL }%
\providecommand \Eprint [0]{\href }%
\providecommand \doibase [0]{http://dx.doi.org/}%
\providecommand \selectlanguage [0]{\@gobble}%
\providecommand \bibinfo  [0]{\@secondoftwo}%
\providecommand \bibfield  [0]{\@secondoftwo}%
\providecommand \translation [1]{[#1]}%
\providecommand \BibitemOpen [0]{}%
\providecommand \bibitemStop [0]{}%
\providecommand \bibitemNoStop [0]{.\EOS\space}%
\providecommand \EOS [0]{\spacefactor3000\relax}%
\providecommand \BibitemShut  [1]{\csname bibitem#1\endcsname}%
\let\auto@bib@innerbib\@empty
\bibitem [{\citenamefont {Aad}\ \emph {et~al.}(2016)\citenamefont {Aad} \emph
  {et~al.}}]{750GeV-ATLAS}%
  \BibitemOpen
  \bibfield  {author} {\bibinfo {author} {\bibfnamefont {G.}~\bibnamefont
  {Aad}} \emph {et~al.} (\bibinfo {collaboration} {ATLAS}),\ }\href@noop {} {\
  (\bibinfo {year} {2016})},\ \bibinfo {note}
  {~ATLAS-CONF-2016-018}\BibitemShut {NoStop}%
\bibitem [{\citenamefont {Chatrchyan}\ \emph {et~al.}(2016)\citenamefont
  {Chatrchyan} \emph {et~al.}}]{CMS:2015dxe}%
  \BibitemOpen
  \bibfield  {author} {\bibinfo {author} {\bibfnamefont {S.}~\bibnamefont
  {Chatrchyan}} \emph {et~al.} (\bibinfo {collaboration} {CMS}),\ }\href@noop
  {} {} (\bibinfo {year} {2016}),\ \bibinfo {note}
  {~CMS-PAS-EXO-16-018}\BibitemShut {NoStop}%
\bibitem [{\citenamefont {Moretti}(2015)}]{Moretti:2014rka}%
  \BibitemOpen
  \bibfield  {author} {\bibinfo {author} {\bibfnamefont {S.}~\bibnamefont
  {Moretti}},\ }\href {\doibase 10.1103/PhysRevD.91.014012} {\bibfield
  {journal} {\bibinfo  {journal} {Phys. Rev.}\ }\textbf {\bibinfo {volume}
  {D91}},\ \bibinfo {pages} {014012} (\bibinfo {year} {2015})}\BibitemShut
  {NoStop}%
\bibitem [{\citenamefont {Landau}(1948)}]{Landau:1948kw}%
  \BibitemOpen
  \bibfield  {author} {\bibinfo {author} {\bibfnamefont {L.~D.}\ \bibnamefont
  {Landau}},\ }\href {\doibase 10.1016/B978-0-08-010586-4.50070-5} {\bibfield
  {journal} {\bibinfo  {journal} {Dokl. Akad. Nauk Ser. Fiz.}\ }\textbf
  {\bibinfo {volume} {60}},\ \bibinfo {pages} {207} (\bibinfo {year}
  {1948})}\BibitemShut {NoStop}%
\bibitem [{\citenamefont {Yang}(1950)}]{Yang:1950rg}%
  \BibitemOpen
  \bibfield  {author} {\bibinfo {author} {\bibfnamefont {C.-N.}\ \bibnamefont
  {Yang}},\ }\href {\doibase 10.1103/PhysRev.77.242} {\bibfield  {journal}
  {\bibinfo  {journal} {Phys. Rev.}\ }\textbf {\bibinfo {volume} {77}},\
  \bibinfo {pages} {242} (\bibinfo {year} {1950})}\BibitemShut {NoStop}%
\bibitem [{\citenamefont {Djouadi}\ \emph {et~al.}(2016)\citenamefont {Djouadi}
  \emph {et~al.}}]{Djouadi:2016eyy}%
  \BibitemOpen
  \bibfield  {author} {\bibinfo {author} {\bibfnamefont {A.}~\bibnamefont
  {Djouadi}} \emph {et~al.},\ }\href {\doibase 10.1007/JHEP03(2016)205}
  {\bibfield  {journal} {\bibinfo  {journal} {JHEP}\ }\textbf {\bibinfo
  {volume} {03}},\ \bibinfo {pages} {205} (\bibinfo {year} {2016})}\BibitemShut
  {NoStop}%
\bibitem [{\citenamefont {Bharucha}\ \emph {et~al.}()\citenamefont {Bharucha}
  \emph {et~al.}}]{Bharucha:2016jyr}%
  \BibitemOpen
  \bibfield  {author} {\bibinfo {author} {\bibfnamefont {A.}~\bibnamefont
  {Bharucha}} \emph {et~al.},\ }\href@noop {} {\ }\Eprint
  {http://arxiv.org/abs/1603.04464} {arXiv:1603.04464 [hep-ph]} \BibitemShut
  {NoStop}%
\bibitem [{\citenamefont {Di~Chiara}\ \emph {et~al.}()\citenamefont {Di~Chiara}
  \emph {et~al.}}]{DiChiara:2016dez}%
  \BibitemOpen
  \bibfield  {author} {\bibinfo {author} {\bibfnamefont {S.}~\bibnamefont
  {Di~Chiara}} \emph {et~al.},\ }\href@noop {} {\ }\Eprint
  {http://arxiv.org/abs/1603.07263} {arXiv:1603.07263 [hep-ph]} \BibitemShut
  {NoStop}%
\bibitem [{\citenamefont {Luo}\ \emph {et~al.}(2016)\citenamefont {Luo} \emph
  {et~al.}}]{Luo:2015yio}%
  \BibitemOpen
  \bibfield  {author} {\bibinfo {author} {\bibfnamefont {M.-x.}\ \bibnamefont
  {Luo}} \emph {et~al.},\ }\href {\doibase 10.1103/PhysRevD.93.055042}
  {\bibfield  {journal} {\bibinfo  {journal} {Phys. Rev.}\ }\textbf {\bibinfo
  {volume} {D93}},\ \bibinfo {pages} {055042} (\bibinfo {year}
  {2016})}\BibitemShut {NoStop}%
\bibitem [{\citenamefont {Han}\ \emph {et~al.}(2016)\citenamefont {Han} \emph
  {et~al.}}]{Han:2016pab}%
  \BibitemOpen
  \bibfield  {author} {\bibinfo {author} {\bibfnamefont {C.}~\bibnamefont
  {Han}} \emph {et~al.},\ }\href {\doibase 10.1007/JHEP04(2016)159} {\bibfield
  {journal} {\bibinfo  {journal} {JHEP}\ }\textbf {\bibinfo {volume} {04}},\
  \bibinfo {pages} {159} (\bibinfo {year} {2016})}\BibitemShut {NoStop}%
\bibitem [{\citenamefont {Kats}\ and\ \citenamefont
  {Strassler}()}]{Kats:2016kuz}%
  \BibitemOpen
  \bibfield  {author} {\bibinfo {author} {\bibfnamefont {Y.}~\bibnamefont
  {Kats}}\ and\ \bibinfo {author} {\bibfnamefont {M.}~\bibnamefont
  {Strassler}},\ }\href@noop {} {\ }\Eprint {http://arxiv.org/abs/1602.08819}
  {arXiv:1602.08819 [hep-ph]} \BibitemShut {NoStop}%
\bibitem [{\citenamefont {Chway}\ \emph {et~al.}()\citenamefont {Chway} \emph
  {et~al.}}]{Chway:2015lzg}%
  \BibitemOpen
  \bibfield  {author} {\bibinfo {author} {\bibfnamefont {D.}~\bibnamefont
  {Chway}} \emph {et~al.},\ }\href@noop {} {\ }\Eprint
  {http://arxiv.org/abs/1512.08221} {arXiv:1512.08221 [hep-ph]} \BibitemShut
  {NoStop}%
\bibitem [{\citenamefont {Furry}(1937)}]{Furry:1937zz}%
  \BibitemOpen
  \bibfield  {author} {\bibinfo {author} {\bibfnamefont {W.~H.}\ \bibnamefont
  {Furry}},\ }\href {\doibase 10.1103/PhysRev.51.125} {\bibfield  {journal}
  {\bibinfo  {journal} {Phys. Rev.}\ }\textbf {\bibinfo {volume} {51}},\
  \bibinfo {pages} {125} (\bibinfo {year} {1937})}\BibitemShut {NoStop}%
\bibitem [{\citenamefont {Dyson}(1949)}]{Dyson:1949ha}%
  \BibitemOpen
  \bibfield  {author} {\bibinfo {author} {\bibfnamefont {F.~J.}\ \bibnamefont
  {Dyson}},\ }\href {\doibase 10.1103/PhysRev.75.1736} {\bibfield  {journal}
  {\bibinfo  {journal} {Phys. Rev.}\ }\textbf {\bibinfo {volume} {75}},\
  \bibinfo {pages} {1736} (\bibinfo {year} {1949})}\BibitemShut {NoStop}%
\bibitem [{\citenamefont {Feynman}(1949)}]{Feynman:1949hz}%
  \BibitemOpen
  \bibfield  {author} {\bibinfo {author} {\bibfnamefont {R.~P.}\ \bibnamefont
  {Feynman}},\ }\href {\doibase 10.1103/PhysRev.76.749} {\bibfield  {journal}
  {\bibinfo  {journal} {Phys. Rev.}\ }\textbf {\bibinfo {volume} {76}},\
  \bibinfo {pages} {749} (\bibinfo {year} {1949})}\BibitemShut {NoStop}%
\bibitem [{\citenamefont {Aad}\ \emph {et~al.}(2012)\citenamefont {Aad} \emph
  {et~al.}}]{Aad:2012tfa}%
  \BibitemOpen
  \bibfield  {author} {\bibinfo {author} {\bibfnamefont {G.}~\bibnamefont
  {Aad}} \emph {et~al.} (\bibinfo {collaboration} {ATLAS}),\ }\href {\doibase
  10.1016/j.physletb.2012.08.020} {\bibfield  {journal} {\bibinfo  {journal}
  {Phys. Lett.}\ }\textbf {\bibinfo {volume} {B716}},\ \bibinfo {pages} {1}
  (\bibinfo {year} {2012})}\BibitemShut {NoStop}%
\bibitem [{\citenamefont {Chatrchyan}\ \emph {et~al.}(2012)\citenamefont
  {Chatrchyan} \emph {et~al.}}]{Chatrchyan:2012xdj}%
  \BibitemOpen
  \bibfield  {author} {\bibinfo {author} {\bibfnamefont {S.}~\bibnamefont
  {Chatrchyan}} \emph {et~al.} (\bibinfo {collaboration} {CMS}),\ }\href
  {\doibase 10.1016/j.physletb.2012.08.021} {\bibfield  {journal} {\bibinfo
  {journal} {Phys. Lett.}\ }\textbf {\bibinfo {volume} {B716}},\ \bibinfo
  {pages} {30} (\bibinfo {year} {2012})}\BibitemShut {NoStop}%
\bibitem [{\citenamefont {Djouadi}\ and\ \citenamefont
  {Lenz}(2012)}]{Djouadi:2012ae}%
  \BibitemOpen
  \bibfield  {author} {\bibinfo {author} {\bibfnamefont {A.}~\bibnamefont
  {Djouadi}}\ and\ \bibinfo {author} {\bibfnamefont {A.}~\bibnamefont {Lenz}},\
  }\href {\doibase 10.1016/j.physletb.2012.07.060} {\bibfield  {journal}
  {\bibinfo  {journal} {Phys. Lett.}\ }\textbf {\bibinfo {volume} {B715}},\
  \bibinfo {pages} {310} (\bibinfo {year} {2012})}\BibitemShut {NoStop}%
\bibitem [{\citenamefont {Eberhardt}\ \emph {et~al.}(2012)\citenamefont
  {Eberhardt} \emph {et~al.}}]{Eberhardt:2012gv}%
  \BibitemOpen
  \bibfield  {author} {\bibinfo {author} {\bibfnamefont {O.}~\bibnamefont
  {Eberhardt}} \emph {et~al.},\ }\href {\doibase
  10.1103/PhysRevLett.109.241802} {\bibfield  {journal} {\bibinfo  {journal}
  {Phys. Rev. Lett.}\ }\textbf {\bibinfo {volume} {109}},\ \bibinfo {pages}
  {241802} (\bibinfo {year} {2012})}\BibitemShut {NoStop}%
\bibitem [{twi({\natexlab{a}})}]{twikiATLAS8TeV}%
  \BibitemOpen
  \href@noop {} {}\bibinfo {howpublished}
  {\url{https://twiki.cern.ch/twiki/bin/view/AtlasPublic/TopPublicResults}}
  ({\natexlab{a}})\BibitemShut {NoStop}%
\bibitem [{twi({\natexlab{b}})}]{twikiATLAS13TeV}%
  \BibitemOpen
  \href@noop {} {}\bibinfo {howpublished}
  {\url{https://twiki.cern.ch/twiki/bin/view/AtlasPublic/ExoticsPublicResults}}
  ({\natexlab{b}})\BibitemShut {NoStop}%
\bibitem [{twi({\natexlab{c}})}]{twikiCMS}%
  \BibitemOpen
  \href@noop {} {}\bibinfo {howpublished}
  {\url{https://twiki.cern.ch/twiki/bin/view/CMSPublic/PhysicsResultsB2G}}
  ({\natexlab{c}})\BibitemShut {NoStop}%
\bibitem [{\citenamefont {Arkani-Hamed}\ \emph {et~al.}(2002)\citenamefont
  {Arkani-Hamed} \emph {et~al.}}]{ArkaniHamed:2002qx}%
  \BibitemOpen
  \bibfield  {author} {\bibinfo {author} {\bibfnamefont {N.}~\bibnamefont
  {Arkani-Hamed}} \emph {et~al.},\ }\href {\doibase
  10.1088/1126-6708/2002/08/021} {\bibfield  {journal} {\bibinfo  {journal}
  {JHEP}\ }\textbf {\bibinfo {volume} {0208}},\ \bibinfo {pages} {021}
  (\bibinfo {year} {2002})}\BibitemShut {NoStop}%
\bibitem [{\citenamefont {Cheng}\ and\ \citenamefont
  {Low}(2003)}]{Cheng:2003ju}%
  \BibitemOpen
  \bibfield  {author} {\bibinfo {author} {\bibfnamefont {H.-C.}\ \bibnamefont
  {Cheng}}\ and\ \bibinfo {author} {\bibfnamefont {I.}~\bibnamefont {Low}},\
  }\href {\doibase 10.1088/1126-6708/2003/09/051} {\bibfield  {journal}
  {\bibinfo  {journal} {JHEP}\ }\textbf {\bibinfo {volume} {09}},\ \bibinfo
  {pages} {051} (\bibinfo {year} {2003})}\BibitemShut {NoStop}%
\bibitem [{\citenamefont {Cheng}\ and\ \citenamefont
  {Low}(2004)}]{Cheng:2004yc}%
  \BibitemOpen
  \bibfield  {author} {\bibinfo {author} {\bibfnamefont {H.-C.}\ \bibnamefont
  {Cheng}}\ and\ \bibinfo {author} {\bibfnamefont {I.}~\bibnamefont {Low}},\
  }\href {\doibase 10.1088/1126-6708/2004/08/061} {\bibfield  {journal}
  {\bibinfo  {journal} {JHEP}\ }\textbf {\bibinfo {volume} {08}},\ \bibinfo
  {pages} {061} (\bibinfo {year} {2004})}\BibitemShut {NoStop}%
\bibitem [{\citenamefont {Low}(2004)}]{Low:2004xc}%
  \BibitemOpen
  \bibfield  {author} {\bibinfo {author} {\bibfnamefont {I.}~\bibnamefont
  {Low}},\ }\href {\doibase 10.1088/1126-6708/2004/10/067} {\bibfield
  {journal} {\bibinfo  {journal} {JHEP}\ }\textbf {\bibinfo {volume} {10}},\
  \bibinfo {pages} {067} (\bibinfo {year} {2004})}\BibitemShut {NoStop}%
\bibitem [{\citenamefont {Hubisz}\ and\ \citenamefont
  {Meade}(2005)}]{Hubisz:2004ft}%
  \BibitemOpen
  \bibfield  {author} {\bibinfo {author} {\bibfnamefont {J.}~\bibnamefont
  {Hubisz}}\ and\ \bibinfo {author} {\bibfnamefont {P.}~\bibnamefont {Meade}},\
  }\href {\doibase 10.1103/PhysRevD.71.035016} {\bibfield  {journal} {\bibinfo
  {journal} {Phys. Rev.}\ }\textbf {\bibinfo {volume} {D71}},\ \bibinfo {pages}
  {035016} (\bibinfo {year} {2005})}\BibitemShut {NoStop}%
\bibitem [{\citenamefont {Cheng}\ \emph {et~al.}(2006)\citenamefont {Cheng}
  \emph {et~al.}}]{Cheng:2005as}%
  \BibitemOpen
  \bibfield  {author} {\bibinfo {author} {\bibfnamefont {H.-C.}\ \bibnamefont
  {Cheng}} \emph {et~al.},\ }\href {\doibase 10.1103/PhysRevD.74.055001}
  {\bibfield  {journal} {\bibinfo  {journal} {Phys. Rev.}\ }\textbf {\bibinfo
  {volume} {D74}},\ \bibinfo {pages} {055001} (\bibinfo {year}
  {2006})}\BibitemShut {NoStop}%
\bibitem [{\citenamefont {Hubisz}\ \emph {et~al.}(2006)\citenamefont {Hubisz}
  \emph {et~al.}}]{Hubisz:2005tx}%
  \BibitemOpen
  \bibfield  {author} {\bibinfo {author} {\bibfnamefont {J.}~\bibnamefont
  {Hubisz}} \emph {et~al.},\ }\href {\doibase 10.1088/1126-6708/2006/01/135}
  {\bibfield  {journal} {\bibinfo  {journal} {JHEP}\ }\textbf {\bibinfo
  {volume} {01}},\ \bibinfo {pages} {135} (\bibinfo {year} {2006})}\BibitemShut
  {NoStop}%
\bibitem [{\citenamefont {Antoniadis}(1990)}]{Antoniadis:1990ew}%
  \BibitemOpen
  \bibfield  {author} {\bibinfo {author} {\bibfnamefont {I.}~\bibnamefont
  {Antoniadis}},\ }\href {\doibase 10.1016/0370-2693(90)90617-F} {\bibfield
  {journal} {\bibinfo  {journal} {Phys.Lett.}\ }\textbf {\bibinfo {volume}
  {B246}},\ \bibinfo {pages} {377} (\bibinfo {year} {1990})}\BibitemShut
  {NoStop}%
\bibitem [{\citenamefont {Appelquist}\ \emph {et~al.}(2001)\citenamefont
  {Appelquist} \emph {et~al.}}]{Appelquist:2000nn}%
  \BibitemOpen
  \bibfield  {author} {\bibinfo {author} {\bibfnamefont {T.}~\bibnamefont
  {Appelquist}} \emph {et~al.},\ }\href {\doibase 10.1103/PhysRevD.64.035002}
  {\bibfield  {journal} {\bibinfo  {journal} {Phys.Rev.}\ }\textbf {\bibinfo
  {volume} {D64}},\ \bibinfo {pages} {035002} (\bibinfo {year}
  {2001})}\BibitemShut {NoStop}%
\bibitem [{\citenamefont {Servant}\ and\ \citenamefont
  {Tait}(2003)}]{Servant:2002aq}%
  \BibitemOpen
  \bibfield  {author} {\bibinfo {author} {\bibfnamefont {G.}~\bibnamefont
  {Servant}}\ and\ \bibinfo {author} {\bibfnamefont {T.~M.~P.}\ \bibnamefont
  {Tait}},\ }\href {\doibase 10.1016/S0550-3213(02)01012-X} {\bibfield
  {journal} {\bibinfo  {journal} {Nucl. Phys.}\ }\textbf {\bibinfo {volume}
  {B650}},\ \bibinfo {pages} {391} (\bibinfo {year} {2003})}\BibitemShut
  {NoStop}%
\bibitem [{\citenamefont {Csaki}\ \emph {et~al.}(2004)\citenamefont {Csaki}
  \emph {et~al.}}]{Csaki:2003sh}%
  \BibitemOpen
  \bibfield  {author} {\bibinfo {author} {\bibfnamefont {C.}~\bibnamefont
  {Csaki}} \emph {et~al.},\ }\href {\doibase 10.1103/PhysRevD.70.015012}
  {\bibfield  {journal} {\bibinfo  {journal} {Phys.Rev.}\ }\textbf {\bibinfo
  {volume} {D70}},\ \bibinfo {pages} {015012} (\bibinfo {year}
  {2004})}\BibitemShut {NoStop}%
\bibitem [{\citenamefont {Cacciapaglia}\ \emph {et~al.}(2010)\citenamefont
  {Cacciapaglia} \emph {et~al.}}]{Cacciapaglia:2009pa}%
  \BibitemOpen
  \bibfield  {author} {\bibinfo {author} {\bibfnamefont {G.}~\bibnamefont
  {Cacciapaglia}} \emph {et~al.},\ }\href {\doibase 10.1007/JHEP03(2010)083}
  {\bibfield  {journal} {\bibinfo  {journal} {JHEP}\ }\textbf {\bibinfo
  {volume} {1003}},\ \bibinfo {pages} {083} (\bibinfo {year}
  {2010})}\BibitemShut {NoStop}%
\bibitem [{\citenamefont {Aaltonen}\ \emph
  {et~al.}(2011{\natexlab{a}})\citenamefont {Aaltonen} \emph
  {et~al.}}]{Aaltonen:2011rr}%
  \BibitemOpen
  \bibfield  {author} {\bibinfo {author} {\bibfnamefont {T.}~\bibnamefont
  {Aaltonen}} \emph {et~al.} (\bibinfo {collaboration} {CDF}),\ }\href
  {\doibase 10.1103/PhysRevLett.106.191801} {\bibfield  {journal} {\bibinfo
  {journal} {Phys. Rev. Lett.}\ }\textbf {\bibinfo {volume} {106}},\ \bibinfo
  {pages} {191801} (\bibinfo {year} {2011}{\natexlab{a}})}\BibitemShut
  {NoStop}%
\bibitem [{\citenamefont {Aaltonen}\ \emph
  {et~al.}(2011{\natexlab{b}})\citenamefont {Aaltonen} \emph
  {et~al.}}]{Aaltonen:2011na}%
  \BibitemOpen
  \bibfield  {author} {\bibinfo {author} {\bibfnamefont {T.}~\bibnamefont
  {Aaltonen}} \emph {et~al.} (\bibinfo {collaboration} {CDF}),\ }\href
  {\doibase 10.1103/PhysRevLett.107.191803} {\bibfield  {journal} {\bibinfo
  {journal} {Phys. Rev. Lett.}\ }\textbf {\bibinfo {volume} {107}},\ \bibinfo
  {pages} {191803} (\bibinfo {year} {2011}{\natexlab{b}})}\BibitemShut
  {NoStop}%
\bibitem [{\citenamefont {Aad}\ \emph {et~al.}()\citenamefont {Aad} \emph
  {et~al.}}]{ATLAS:2011mda}%
  \BibitemOpen
  \bibfield  {author} {\bibinfo {author} {\bibfnamefont {G.}~\bibnamefont
  {Aad}} \emph {et~al.} (\bibinfo {collaboration} {ATLAS}),\ }\href@noop {} {\
  }\bibinfo {note} {ATLAS-CONF-2011-036}\BibitemShut {NoStop}%
\bibitem [{\citenamefont {Chatrchyan}\ \emph {et~al.}()\citenamefont
  {Chatrchyan} \emph {et~al.}}]{CMS:2012dwa}%
  \BibitemOpen
  \bibfield  {author} {\bibinfo {author} {\bibfnamefont {S.}~\bibnamefont
  {Chatrchyan}} \emph {et~al.} (\bibinfo {collaboration} {CMS}),\ }\href@noop
  {} {\ }\bibinfo {note} {CMS-PAS-SUS-12-009}\BibitemShut {NoStop}%
\bibitem [{\citenamefont {Stelzer}\ and\ \citenamefont
  {Long}(1994)}]{Stelzer:1994ta}%
  \BibitemOpen
  \bibfield  {author} {\bibinfo {author} {\bibfnamefont {T.}~\bibnamefont
  {Stelzer}}\ and\ \bibinfo {author} {\bibfnamefont {W.~F.}\ \bibnamefont
  {Long}},\ }\href {\doibase 10.1016/0010-4655(94)90084-1} {\bibfield
  {journal} {\bibinfo  {journal} {Comput. Phys. Commun.}\ }\textbf {\bibinfo
  {volume} {81}},\ \bibinfo {pages} {357} (\bibinfo {year} {1994})},\ \Eprint
  {http://arxiv.org/abs/hep-ph/9401258} {arXiv:hep-ph/9401258 [hep-ph]}
  \BibitemShut {NoStop}%
\bibitem [{\citenamefont {Lai}\ \emph {et~al.}(2000)\citenamefont {Lai} \emph
  {et~al.}}]{Lai:1999wy}%
  \BibitemOpen
  \bibfield  {author} {\bibinfo {author} {\bibfnamefont {H.~L.}\ \bibnamefont
  {Lai}} \emph {et~al.} (\bibinfo {collaboration} {CTEQ}),\ }\href {\doibase
  10.1007/s100529900196} {\bibfield  {journal} {\bibinfo  {journal} {Eur. Phys.
  J.}\ }\textbf {\bibinfo {volume} {C12}},\ \bibinfo {pages} {375} (\bibinfo
  {year} {2000})}\BibitemShut {NoStop}%
\bibitem [{\citenamefont {Sato}\ and\ \citenamefont
  {Tobioka}()}]{Sato:2016hls}%
  \BibitemOpen
  \bibfield  {author} {\bibinfo {author} {\bibfnamefont {R.}~\bibnamefont
  {Sato}}\ and\ \bibinfo {author} {\bibfnamefont {K.}~\bibnamefont {Tobioka}},\
  }\href@noop {} {\ }\Eprint {http://arxiv.org/abs/1605.05366}
  {arXiv:1605.05366 [hep-ph]} \BibitemShut {NoStop}%
\bibitem [{\citenamefont {No}()}]{No:2016htu}%
  \BibitemOpen
  \bibfield  {author} {\bibinfo {author} {\bibfnamefont {J.~M.}\ \bibnamefont
  {No}},\ }\href@noop {} {\ }\Eprint {http://arxiv.org/abs/1605.05900}
  {arXiv:1605.05900 [hep-ph]} \BibitemShut {NoStop}%
\end{thebibliography}%
\end{document}